\documentclass[sn-mathphys-num]{sn-jnl}

\usepackage{algorithm2e}
\usepackage{float}
\usepackage{comment}
\usepackage{multirow}

\usepackage{graphicx}%
\usepackage{multirow}%
\usepackage{amsmath,amssymb,amsfonts}%
\usepackage{amsthm}%
\usepackage{mathrsfs}%
\usepackage[title]{appendix}%
\usepackage{xcolor}%
\usepackage{textcomp}%
\usepackage{manyfoot}%
\usepackage{booktabs}%
\usepackage{algorithmicx}%
\usepackage{algpseudocode}%
\usepackage{listings}%
\usepackage{array}
\usepackage{booktabs}
\usepackage{multirow}

\theoremstyle{thmstyleone}%
\theoremstyle{thmstyletwo}%

\theoremstyle{thmstylethree}%

\begin{document}
		\title{Complete Key Recovery of a DNA-based Encryption and Developing a Novel Stream Cipher for Color Image Encryption: Bio-SNOW}

\author*[1]{\fnm{Yash} \sur{Makwana}}\email{yash9281@gmail.com}

\author[1]{\fnm{Anupama} \sur{Panigrahi}}\email{anupama.panigrahi@gmail.com}
\equalcont{These authors contributed equally to this work.}

\author[2]{\fnm{S.K.} \sur{Pal}}\email{skptech@yahoo.com}
\equalcont{These authors contributed equally to this work.}

\affil*[1]{\orgdiv{Department of Mathematics}, \orgname{University of Delhi}, \orgaddress{ \city{New Delhi}, \postcode{110007}, \state{New Delhi}, \country{India}}}

\affil[2]{\orgdiv{Scientific Analysis Group(SAG)}, \orgname{DRDO}, \orgaddress{\street{MetCalfe House}, \city{New Delhi}, \postcode{110054}, \state{New Delhi}, \country{India}}}

		\abstract{Recent studies have explored DNA-based algorithms for IoT security and image encryption. A similar encryption algorithm was proposed by Al-Husainy et. al. in 2021 Recently, Al-Husainy et al.in 2021,  proposed an encryption algorithm based on DNA processes for Internet of Things(IoT) applications. Upon finding low avalanche effect in our experiments, we first report related-key attack  and later propose a key recovery attack on full cipher, which generates the complete key with just two plaintext-ciphertext blocks with time complexity of O(1). Upon discovering these serious weaknesses, we improve the security of encryption algorithm against above reported attacks  by employing a bio-inspired SBOX and adding some tweaks in the cipher. A lot of research has been done in developing image encryption algorithm based on DNA and chaotic maps.  Inspired from the design of SNOW-3G, a well known cipher, primarily used in mobile communications and considering DNA-based functions as building blocks, we propose a new DNA-based stream cipher-Bio-SNOW. We discuss its security in response to various kinds of attacks and find that it passes all NIST randomness tests. We also find that the speed of Bio-SNOW is around twice that of SNOW-3G. Moreover, by means of histogram and correlation analysis, we find that Bio-SNOW offers robust image encryption.  These results highlight Bio-SNOW as a promising DNA-based cipher for lightweight and image cryptography applications. 
	}


\maketitle
 \vspace{0.5cm}

\section{Introduction}

Historically, cryptography focused solely on securing data through simple operations that concealed information from intruders. Today, its scope has expanded to include multi-party computation, blockchain, cryptocurrency, hash functions, and digital signatures, addressing not only data confidentiality but also authentication and hidden computations. Cryptography can be broadly classified into symmetric and asymmetric types. Symmetric cryptography uses one key for both encryption and decryption, while asymmetric cryptography employs two keys—one for encryption and one for decryption—enabling applications like authentication, secure communication, and multi-party computation.
 
In recent trends, Deoxyribonucleic acid(DNA) has been extensively used to define cryptosystems for general as well as image and video encryption. The chaotic maps provides a good source of randomness and thus used in such types of cryptsystems \cite{alawida2024novel, alawida2024enhancing}. Some research works applies these maps and DNA encoding for color image encryption, which can be improved further by considering $4D$ chaotic maps \cite{masood2022new,wang2022chaotic}. However, these types of systems solves only confidentiality problems and authentication is still left to be ensured. A complete color image encryption system along with authentication protocol is proposed in \cite{JASRA2022117861}.

The Internet of Things (IoT) represents a network of interconnected smart devices that autonomously generate and exchange data in real time, eliminating the need for direct human involvement. However, the security landscape of IoT is reminiscent of early computer systems, where security concerns were often overlooked during the design phase \cite{}. Consequently, each IoT system becomes vulnerable to exploitation by malicious actors. Numerous approaches and methodologies have been proposed by both academic and industrial researchers to mitigate such risks \cite{al2023modified, prakasam2021enhanced} .

Among these, DNA computing emerges as a novel and burgeoning field dedicated to addressing data encryption challenges within the IoT framework. This innovative technique leverages concepts from DNA computing to facilitate rapid and secure data transfer among connected devices, all while maintaining a commitment to low power consumption. As the IoT ecosystem continues to expand, integrating advanced security measures becomes imperative to safeguard against potential threats and ensure the robustness of IoT-enabled applications.

Traditional encryption techniques like Advanced Encryption Standard (AES), hashing, Rivest– Shamir-Adleman (RSA), and Elliptic Curve Cryptography (ECC) are ill-suited for IoT devices due to their resource constraints, including limited power, memory, and processing capabilities (\cite{sepranos2019challenges}). Consequently, the exploration of lightweight cryptography has gained traction with a lot of ciphers proposed. (\cite{buchanan2017lightweight}). However, implementing these lightweight techniques poses challenges in balancing memory, speed, power consumption, and maintaining robust data security.

Al-Husainy et al. \cite{al2021lightweight} devised an algorithm for IoT applications keeping the design as simple as possible whose details are in section \ref{oldcipher}. However, due to the simple structure there is a serious vulnerability which gives us a complete attack on this cipher. Moreover, while analyzing it was found that this algorithm is vulnerable to related key and distinguishing attacks due to weak avalanche effect.  In section \ref{attacksonold}, we describe several attacks which includes the complete break. Later, we propose an improvement to this cipher where we use an Amino Acids S-Box for high avalanche effect and include mutator vector in encryption process against the vulnerabilities. In section \ref{practicalresults}, we analyze PSNR, Entropy, and Avalanche effect and compare it with the old cipher.

There is a need for efficient ciphers in the field of communication and SNOW-3G is a cipher of the SNOW family that is used in 3G and 4G communications (\cite{ekdahl2018new}, \cite{ekdahl2000snow}). In section \ref{newcipher}, inspired by SNOW we extend cipher from (\cite{sadeg2010encryption}) and propose a new secure and efficient algorithm- Bio-SNOW. We present some tests in comparison to the SNOW cipher for our claims. 

Recent advancements in image encryption have introduced DNA-based algorithms, with many of these relying on chaos-based encryption techniques (\cite{CHAI2017197,Malik,Samiullah}). Chaos-based image ciphers are recognized for their ability to achieve low correlation and high entropy in encrypted images (\cite{FU20131000, li2022novel, lai2023image, yan2023color}). However, several vulnerabilities have been identified in such ciphers (\cite{su2017cryptanalysis, CHEN201569}). Beyond chaos-based methods, numerous other designs for image encryption have been proposed and subsequently subjected to cryptanalysis \cite{ye2024visual}. We present the application of cipher Bio-SNOW in the realm of image cryptography and discuss the essential security measures required for robust image encryption algorithms.

\section{Preliminaries}
While DNA-based encryption algorithms have been proposed in the literature, a limited body of research delves into simulating the intricate processes of the central dogma. Notably, we have identified only a few pertinent works in this domain. One such instance is a straightforward cryptographic method, inspired by analogous principles and detailed in (\cite{ning2009pseudo}). Additionally, Amin et al. in 2006 \cite{amin2006dna}, introduced a symmetric encryption algorithm named YAEA. This algorithm transforms binary data, including plaintext messages and images, into sequences of DNA nucleotides. Subsequently, efficient search algorithms are employed to identify multiple occurrences of a four-DNA-nucleotide sequence representing a binary octet plaintext character within a Canis Familiaris genomic chromosome. Randomly selected pointers of these four DNA nucleotides for each plaintext character are consolidated into a file, constituting the ciphertext. Remarkably, this approach can be employed to reinforce established cryptographic algorithms.

The central dogma of molecular biology, elucidated by Watson and Crick \cite{watson1953molecular} in a seminal paper published in Nature, posits that information flow from protein to either protein or nucleic acid is inherently restricted. Furthermore, it elucidates the fundamental principles governing the standard biological information flow, encompassing DNA to DNA (DNA replication), DNA to RNA (transcription), and RNA to protein (translation).

\paragraph{FSR:}
A Feedback shift register is a array of where each cell except the one at the end. The value of the first cell is updated according to a function of the values of another cell. For example if the cells are given by $s_0(t), s_1(t), s_2(t) \dots s_n(t+1)$ then,  
\begin{eqnarray*}
	s_1(t+1)&=&s_0(t) \\
	s_2(t+1)&=&s_1(t) \\
	s_3(t+1) &=& s_2(t) \\
	.      & &  \\
	.      & &   \\
	s_n(t+1) &=& s_{n-1}(t) \\
	s_0(t+1)&=& f(s_0(t), s_1(t)), \dots s_n(t) 
\end{eqnarray*}
\paragraph{FSM:} Finite State Machine can have different states. Unlike FSR, it does not  have a fixed shifting mechanism. Its state can take any possible values prescribed. For example,  if $S$ is the current state of FSM, then $S$ can be updated by applying AES Round function on $S$. Then the new state will be:\\ \[ S(t+1)= AES(S) \]
\\

In this paper, we try to solve the problem of security of DNA-based ciphers by proposing an attack on a DNA-based encryption scheme and introducing new cipher in this regard. The rest of the paper is organized as follows: section \ref{oldcipher} discusses several attacks including complete key recovery on a DNA-based cipher which was introduced in \cite{al2021lightweight}, and proposes some improvements to withstand such type of attacks. In section \ref{newcipher}, we propose a new cipher BioSNOW by using FSRs and Bio-inspired functions. Section \ref{securityandimage} discusses its security properties as well as application to image cryptography.  Finally, section \ref{sec:conclusion} concludes the paper with a summary of our findings and a discussion of potential future research directions.

\section{DNA based cipher and its break}\label{oldcipher}
\subsection{Specifications}
Al-Husainy et al.\cite{al2021lightweight} introduced a novel cipher harnessing the inherent randomness found within DNA. The researchers opted for a straightforward structure to adapt it for lightweight cryptography. However, this apparent simplicity led to a critical vulnerability in the cipher's design. The proposed cipher formulation is outlined as follows:
\begin{enumerate}
	\item \textbf{Converting into Blocks} We partition the plaintext into blocks of size $(n \times 8) \times (n \times 8)$. The parameter $n$ delineates the cipher's specific configurations. For ease of representation, we set $N=n\times8$.
	\item \textbf{Key Selection}
	A wealth of approximately $163$ million publicly available DNA sequences serves as our source of randomness. Leveraging this, we extract three segments of the secret key from these sequences, converting each DNA term into bits using a specific coding method as given in \ref{fig:encoding}(b). This results in the formulation:
	\[K=(RV,CV,MV)\]
	where \textbf{RV}=Row Vector=$(r_1, r_2, \dots r_{N})$, \textbf{CV}= Column Vector=$(c_1, c_2, \dots c_{N})$, and \textbf{MV}= Mutator Vector= $(mv_1, mv_2, \dots mv_{N})$. Each component comprises $N$ bits, culminating in a total key size of $(3 \times 8 \times n)= (24 \times n)$.
	\begin{figure}[h]\label{fig:encoding}
		\centering
		\includegraphics[scale=0.42]{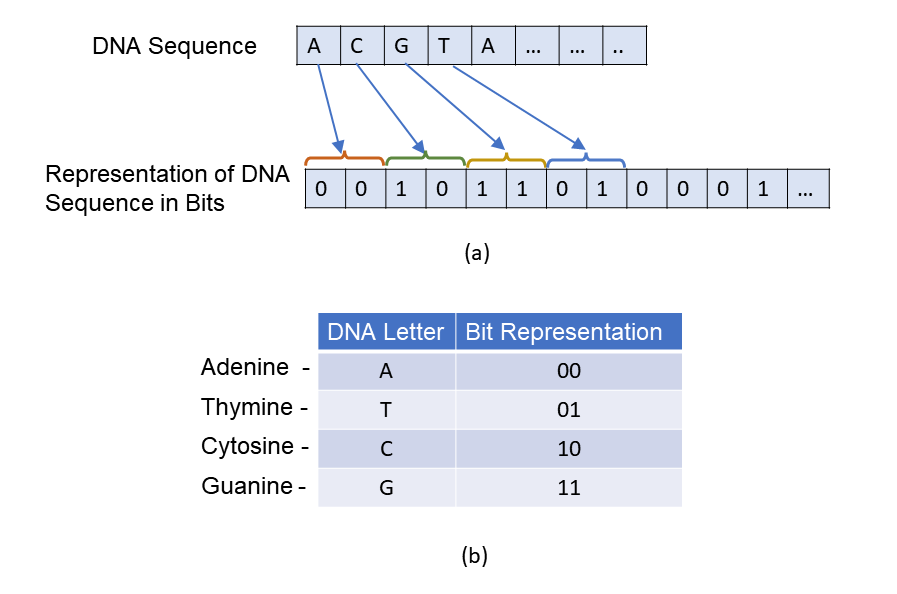}
		\caption{(a)Key Selection and (b) Encoding scheme (\cite{al2021lightweight})}
	\end{figure}
	\item \textbf{Encryption Process}
	\begin{enumerate}
		\item \textbf{Substitution Phase:} Each row and column of the SBlock undergoes an XOR operation with RV and CV respectively, leading to the following equations:
		\begin{eqnarray*}
			p_{ij}'=p_{ij} \oplus r_i \oplus c_i
		\end{eqnarray*}
		where $i, j \in \{ 1, \dots N \}$.\\
		For each SBlock, the RV and CV vary according to the update mechanism utilizing MV.
		\begin{figure}[h]
			\centering
			\includegraphics[scale=0.43]{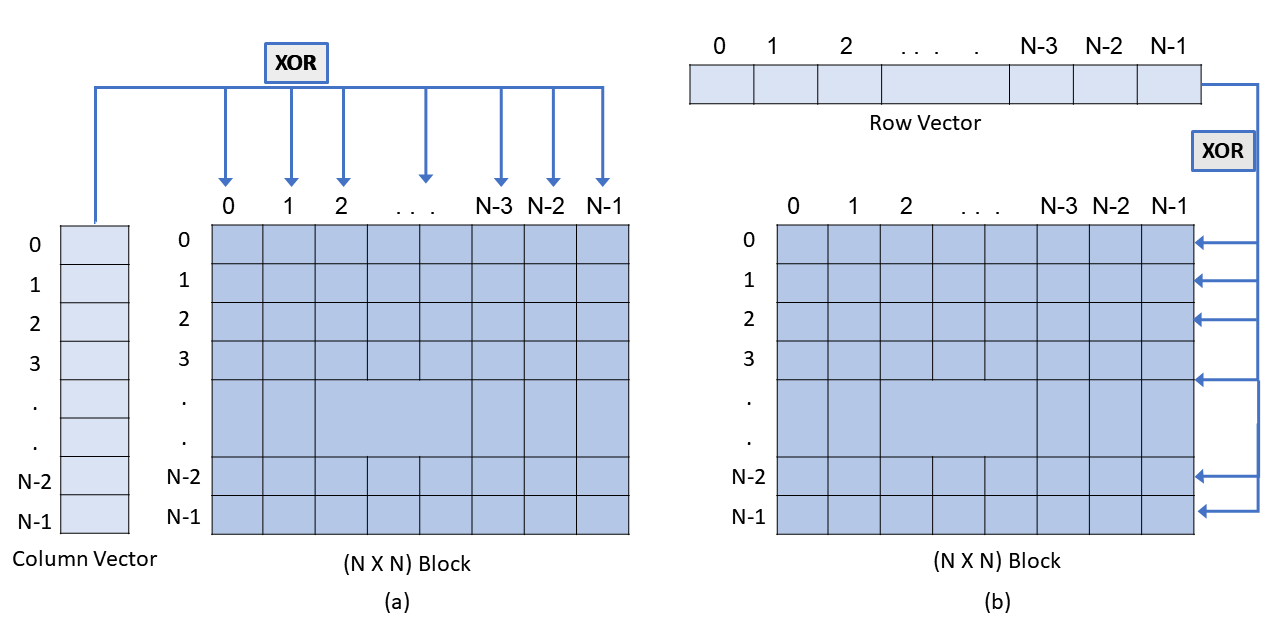}
			\caption{XOR of row and column vector with plaintext block (\cite{al2021lightweight})}
		\end{figure}
		\item \textbf{Transposition Operation Stage}
		This stage is governed by the algorithm \ref{alg:transposition} that is described in Figure \ref{fig:Transposition} \\
		\begin{algorithm}[H]
			\SetAlgoLined
			\KwData{$p_{ij}'$}
			
			\For{$i=1$ to $N$}{
				\If{$r_i \oplus c_i=1$}{
					Interchange $i^{th}$ row and column
				}
			}
			\caption{Transposition}
			\label{alg:transposition}
		\end{algorithm}
		\begin{figure}[h]
			\centering
			\includegraphics[width=0.4\textwidth]{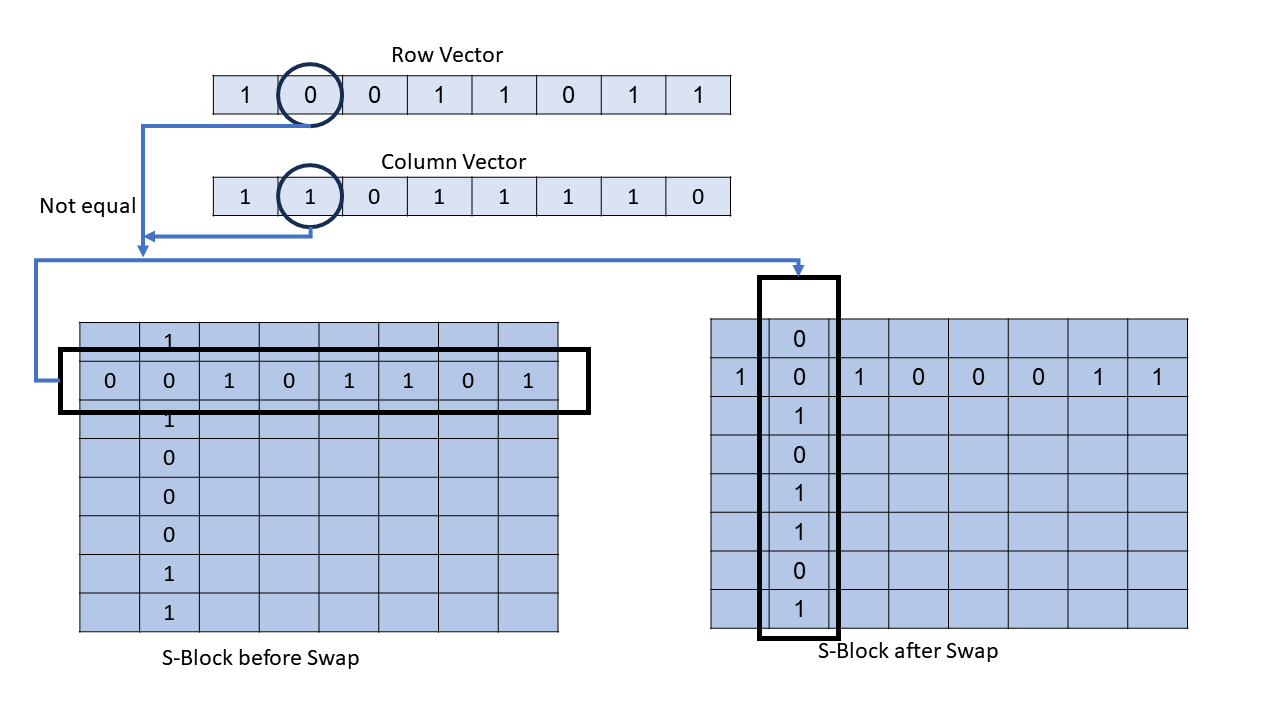}
			\caption{Transposition Phase (\cite{al2021lightweight})}
			\label{fig:Transposition}
		\end{figure}
		\item \textbf{Key Generation Phase}
		
		Distinct RV and CV values are allocated for each SBlock, determined through the mechanism given in Figure \ref{keygeneration}.	
		\begin{figure}[h] 
			\centering
			\includegraphics[scale=0.35]{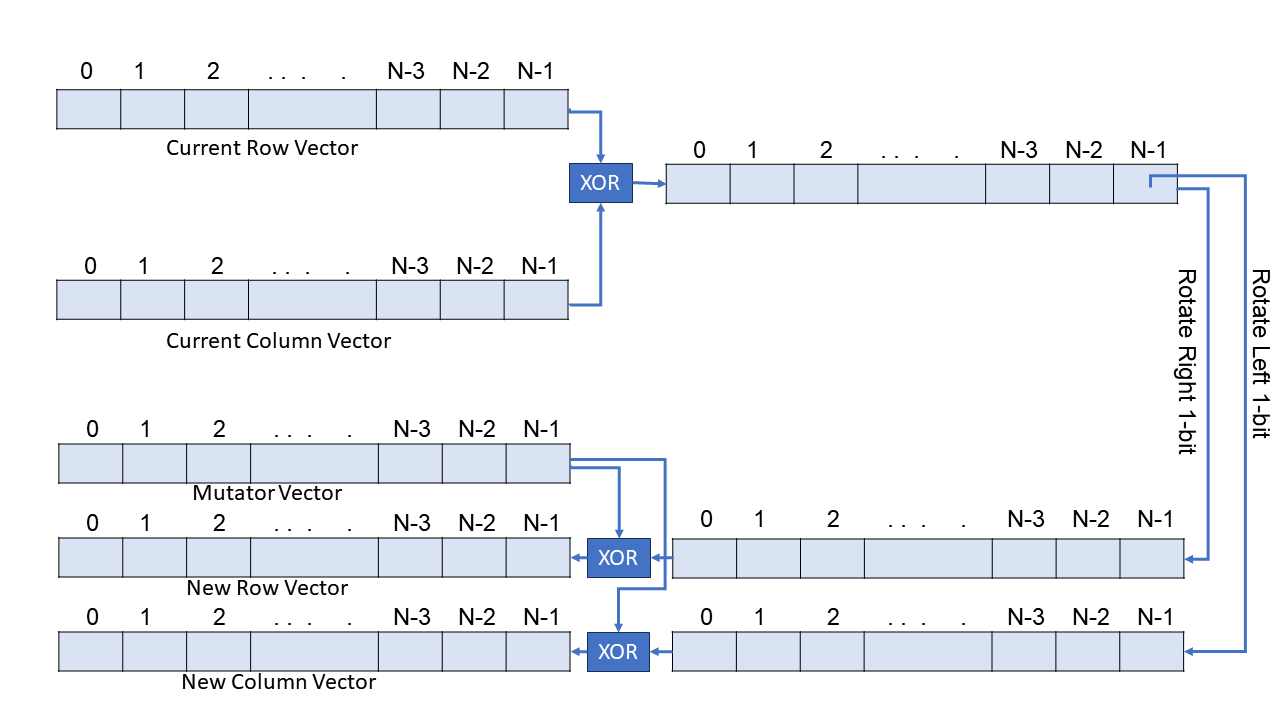}
			\caption{Key Generation Phase (\cite{al2021lightweight})}
			\label{keygeneration}
		\end{figure}
	\end{enumerate}
\end{enumerate}

\subsection{Related-Key Attacks} \label{attacksonold} To prevent the cipher from related key attacks, it is necessary for the cipher to have avalanche effect, that is there is a big change in ciphertext if only a minor update is done in the key. It is given by equation \eqref{eq:avalanche}. 
\begin{equation} \label{eq:avalanche}
	\text{Avalanche Effect}= \frac{\text{No. of bits changed after new key is used}}{\text{Total number of bits in data}}
\end{equation}
When we checked the Avalanche effect of the old cipher, the results were divergent from the one published in (\cite{al2021lightweight}). 
\begin{table}[h]
	\centering
	\begin{tabular}{ccc}
		\toprule Sr. No. & Data Size & Avalanche Effect for 1 bit
		\\ \midrule
		1 & 49152 & 34.53\% \\ 
		2 & 709200		& 31.11 \%\\ 
		3 & 889,200	& 33.81 \%\\ 
		4 & 950,878 &  36.71 \% \\ \bottomrule
	\end{tabular}
	\caption{Avalanche Effect of Old Cipher}
	\label{table:avalancheoldcipher}
\end{table}
Table \ref{table:avalancheoldcipher}  shows that changing one bit of key keeps around $68\%$ of bits exactly same. In many scenarios where key is updated by successively increasing the value and generating two ciphertexts for same plaintext, this can pose a serious threat.   
\subsection{Complete break of cipher}
\subsubsection{Keysize reduction Attack}
We saw that the key has size of $3\times N$ bits, which means that brute force attack will require $2^{3N}$ iterations. We can reduce the effective key size from $3N$ to $N$. This is possible as mutator vector does not play any role in encryption of one SBlock and we shall show in section \ref{sec:mutator vector} that one can derive mutator vector if RV and CV are known for two consecutive SBlocks. \\

After substitution phase, we have following relations.
\begin{eqnarray*}
	p_{ij}'=p_{ij} \oplus r_i \oplus c_i
\end{eqnarray*}
After transposition phase, depending on the mutator vector bits, the $p_{ij}'$ and $p_{ji}'$ may be interchanged. But irrespective of mutator bits, the diagonal elements are not changed. let $q_{ij}$ be the final ciphertext, then we have
\begin{equation}
	q_{ii}= p_{ii} \oplus r_i \oplus c_i
\end{equation}

Considering known plaintext-ciphertext attack, we may know some of these bit pairs and only from $N$ bit pairs of $(p_{ii}, q_{ii})$ we have $N$ equations for these diagonal elements which give us:
\begin{equation}\label{rici}
	r_i \oplus c_i = p_{ii} \oplus q_{ii}
\end{equation}
Thus, the complexity of brute force is reduced from $2^{3N}$ to $2^{N}$. This key size reduction introduces restrictions on choice of parameter $N$.

\subsubsection{Recovering row and column vector} \label{sec:complete break}
We can improve this attack to completely break the cipher by employing some hidden vulnerabilities.  From equation (\ref{rici}), we know the exact value of $r_i \oplus c_i$. We also saw that this is the same value that defines the transposition phase. If its value is $1$ the corresponding row and columns are interchanged. Hence, we know whether $p_{ij} \oplus r_{i} \oplus c_{j}$ equals $q_{ij}$ or $q_{ji}$. 
We have already reduced the key size from $3N$ to $N$ and now we have $N$ unknowns. To get exact values of these $N$ variables, we need only $N$ linearly independent equations. Considering one SBlock, we have $N^2-N$ such equations. From the previous section and these relations, we can find the exact value of row and column vectors. 
\subsubsection{Recovering mutator vector} \label{sec:mutator vector}
We observe that the mutator vector is same for all SBlocks, and its only function is to update the row and column vector. By knowing the RV and CV of two consecutive SBlocks, we can easily find the MV. \\

Let $r_i'$ and $c_i'$ be the new row and column vector bits respectively. From the steps of Key generation algorithm, we have:

\begin{algorithm}[H]
	\SetAlgoLined
	\KwData{Input sequences $r_i$ and $c_i$, mutation values $mv_i$}
	\KwResult{Updated sequences $r_i'$ and $c_i'$}
	\For{$i=1$ \KwTo $N$}{
		Calculate $s_i \leftarrow r_i \oplus c_i$\;
	}
	$ls_N \leftarrow s_1$; \\
	\For{$i \leftarrow 1$ \KwTo $N-1$}{
		$ls_{i} \leftarrow s_{i+1}$\;
	}
	\For{$i \leftarrow 2$ \KwTo $N$}{
		$rs_{i} \leftarrow s_{i-1}$\;
	}
	$rs_1 \leftarrow s_N$
	\For{$i \leftarrow 1$ \KwTo $N$}{
		$r_i' \leftarrow ls_i \oplus mv_i$\;
		$c_i' \leftarrow rs_i \oplus mv_i$\;
	}
	\caption{Sequence Update Algorithm}
\end{algorithm}
\vspace{.5cm}

If $r_i, c_i$ are known, we can reach to the equation 
\begin{eqnarray}
	r_i' &=& ls_i \oplus mv_i \nonumber \\ 
	c_i' &=& rs_i \oplus mv_i \label{mveqn}
\end{eqnarray}
where $ls_i$ and $rs_i$ are known. \\

Now, if $r_i'$ and $c_i'$ are also known as discussed in section \ref{sec:complete break}, we can easily find the values of $mv_i$ from equations (\ref{mveqn}).Thus by knowledge of only $2$ SBlocks we can break the complete cipher and extract whole plaintext. We see that the time complexity is $O((2N)^3)$ considering solving linear equations in $2N$ variables. As $N$ is constant for the cipher, the complexity becomes $O(1)$.

\subsection{Getting rid of these vulnerability}
Most of the above attacks exploit the vulnerability that after transposition the diagonal elements are left unchanged. We provide an improvement to tackle this specific vulnerability by making use of mutator vector in transposition phase.\vspace{0.4cm}

\textbf{Proposed Improvement}

After performing the transpose, we add one more step of XOR as:

\[ q_{ii}'= q_{ii} \oplus mv_i \]

\begin{figure}[h]
	\centering
	\includegraphics[scale=0.42]{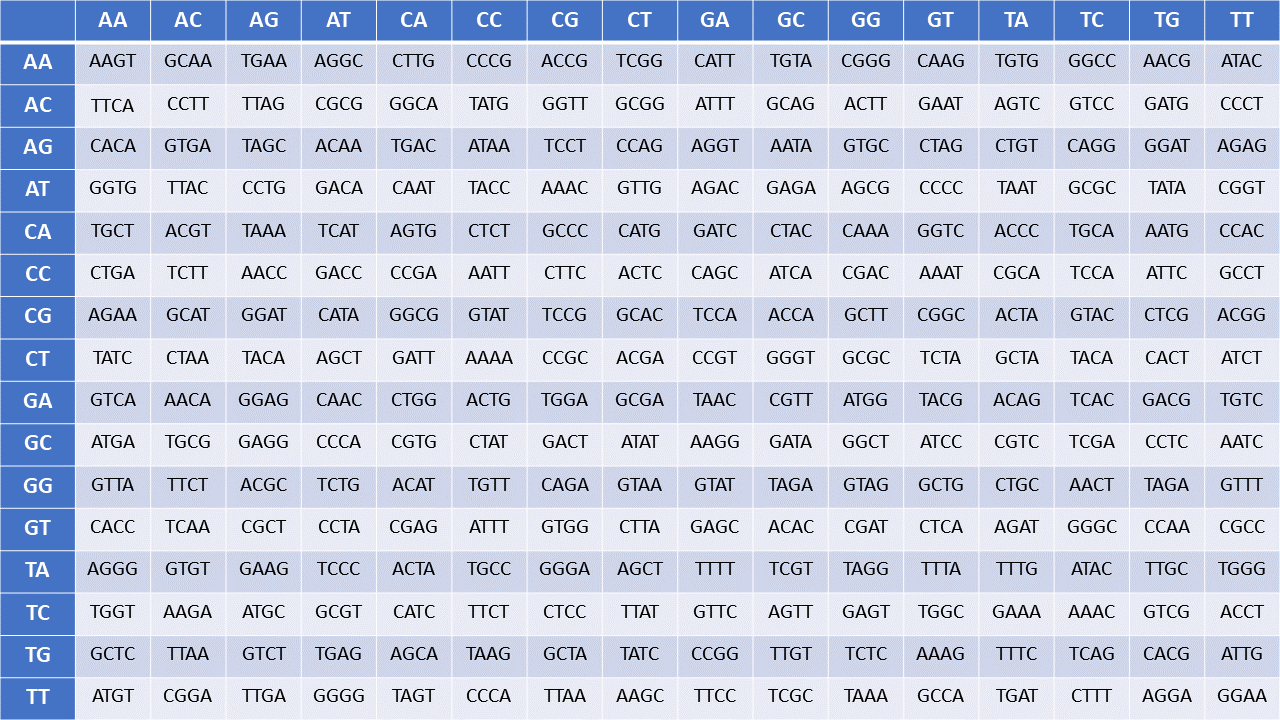}
	\caption{Amino Acid Table (\cite{sadeg2010encryption})}
	\label{fig:AminoAcid}
\end{figure}

This one step makes most of the above attacks a failure as now we don't have any exact relations which give value of $r_i \oplus c_i$. This value was exploited to give us a complete break of the cipher. \\
Now, we have to improve this encryption such that we have a good value of avalanche effect. For this purpose, we include use of an SBOX in the beginning of encryption which is inspired from Amino Acid Table in Figure \ref{fig:AminoAcid}. \\
\vspace{.5cm}
 So, the improved algorithm is:

\begin{algorithm}[H]
	
	\SetAlgoLined
	\KwData{Plaintext $p_i$, keys $r_i, c_i, m_i$}
	\KwResult{Ciphertext $q_i$}
	Create Blocks from $p_i$, naming as \textbf{Blocks[i]} for i=0 \KwTo N-1
	
	\tcc{where $N$ is number of blocks.}
	\textbf{Initialize} keys as rv=r, cv=c, mv=m  
	
	\tcc{For each Block}
	\For{$i=0$ \KwTo $N-1$}{
		\If{$i>0$}{Apply \textbf{KeyVariation}(rv,cv,mv) as defined}

		\For{j=0 \KwTo 7}{
			\For{k=0 \KwTo 7}{
				\textbf{Blocks}[i][j][k]= \textbf{Blocks}[i][j][k] $\oplus rv_j \oplus cv_k$
			}
		}
		
		\textbf{Blocks}[i]= AminoAcidBox(\textbf{Blocks}[i])
		
		\tcc{Applying AminoAcid SBox that is defined in Figure \ref{fig:AminoAcid}}
		
		\For{$k=0$ \KwTo $7$}{
			\If{$cv_k==rv_k$}{
				\For{j=0 \KwTo 7}{(\textbf{Blocks}[i][j][k], \textbf{Blocks}[i][k][j])= (\textbf{Blocks}[i][k][j], \textbf{Blocks}[i][j][k])}}	
			
			\textbf{Blocks}[i][k][k]=Blocks[i][k][k] $\oplus mv_k$
		}
		
	}	
	Create ciphertext $q_i$ from these output Blocks.

	\KwOut{$q_i$}
	\caption{Improved Encryption Algorithm}
\end{algorithm}
\vspace{.5cm}
By employing above techniques we have made the cipher secure against the related-key and the complete key recovery attack.

\subsection{Practical results} \label{practicalresults}
Peak Signal to Noise Ratio(PSNR) value, Avalanche effect  and Entropy are analyzed to determine quality of ciphertext produced by encryption algorithms. We find that the new cipher has Avalanche effect of around $50\%$ compared to just $32\%$(average) of old one. 
\subsubsection{PSNR Value}A good cryptosystem results in high confusion and diffusion in source data and these effects can be measured numeriacally using PSNR given by equation \eqref{eq:psnr1} and \eqref{eq:psnr2} that were given in (\cite{al2021lightweight}). 
\begin{equation}\label{eq:psnr1}
	\text{NMAE}=\frac{\sum_{i=1}^{\text{SSize}-1} |S(k)-E(k)|}{\text{SSize}} \times 100
\end{equation}

\begin{equation}\label{eq:psnr2}
	\text{PSNR}_{db}= 10. \log_{10} \left( \frac{\text{Max}^2_I}{NMAE} \right)
\end{equation}
where $\text{Max}^2_I$ is the maximum possible byte value  of the data $S$  and \textit{db} refers to decibel. \\

The PSNR values for old and improved cipher are given below for different data sizes where size of key is $24$ bits and plaintext is in multiples of $64$ for easy implementation.

\begin{table}[h]
	\centering
	\begin{tabular}{cccc}
		\toprule	Sr. No. & Size of data(in bytes) & PSNR (old) & PSNR (improved) \\ \midrule
		1	& 49,152 & 8.859537958376132& 8.821085396907058 \\ 
		2 & 709,200 & 8.660211999594875 & 8.800436864642574 \\ 
		3 & 889,200 &  9.562951822070005  & 8.85770494223293  \\ 
		4 & 950,878 & 9.113875223755075 & 8.835434306613376\\ \bottomrule
	\end{tabular}
	\caption{PSNR values of Improved cipher}
\end{table}

\begin{figure}[h]
	\centering
	\includegraphics[scale=1.3]{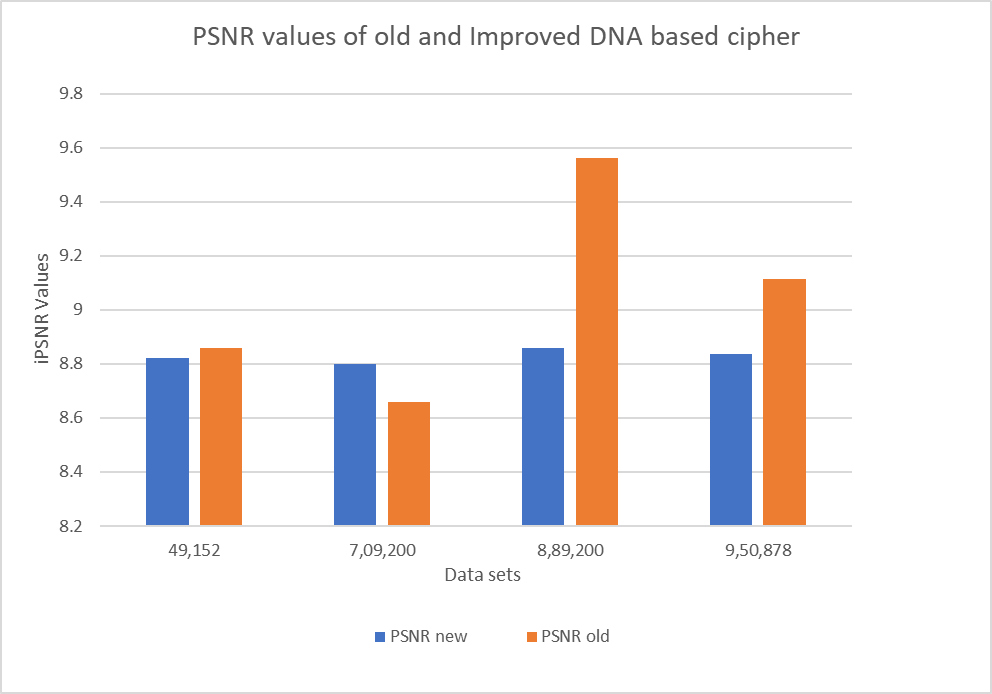}
	\caption{PSNR values of old and improved cipher}
\end{figure}

\subsubsection{Entropy Value}
Entropy can also be used to analyze the quality of encryption as it is difficult to predict the actual content of data whose information entropy is high enough. It can be calculated by equation \eqref{entropyeq}
\begin{equation} \label{entropyeq}
	\text{Entropy}= -\sum_{i=1}^{n} P_i \log_2(P_i)
\end{equation}
where $n$ is the number of different data values and $P_i$ is the probability of occuring the data value. The entropy for different data sizes are given as below: 
\begin{table}[h]
	\centering
	\begin{tabular}{cccc}
		\toprule Sr. No. & Data Size(in bytes) & Entropy (old) & entropy(new) \\ \midrule 
		1	& 49,152 & 7.996218885269518 & 7.992868559137702 \\ 
		2 & 709,200 & 7.999742044278854 & 7.992853569955033 \\ 
		3 & 889,200 & 7.999794464415016 & 7.973759874012821 \\
		4 & 950,878 &  7.999811652339861  &  7.995461274986118  \\ \bottomrule
	\end{tabular}
	\caption{Entropy of Improved Cipher}
\end{table}
\vspace{0.3cm}
\begin{figure}[h]
	\centering
	\includegraphics[scale=1.3]{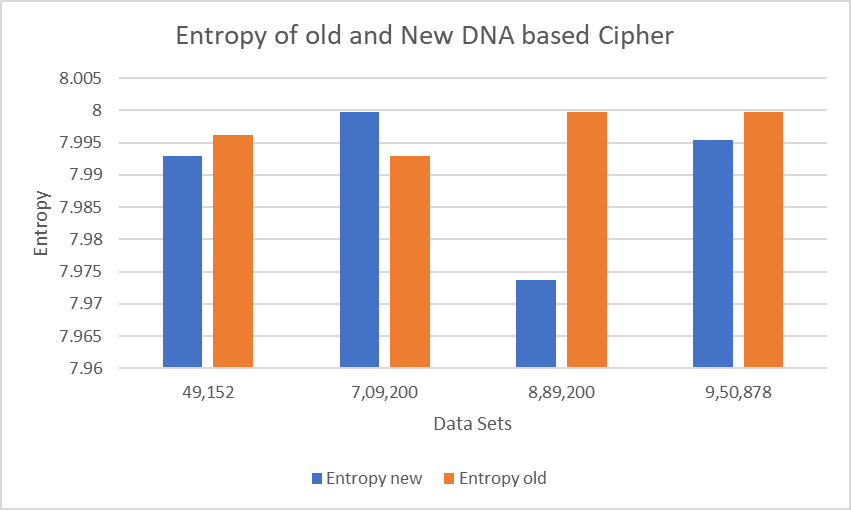}
	\caption{Entropy of Old and Improved Cipher}
\end{figure}
\subsubsection{Avalanche Effect}
We compare the results for the cipher given in (\cite{al2021lightweight}) with our improved cipher after taking average over $24$ observations, one for each bit of the key.
\begin{table}[h]
	\centering
	\begin{tabular}{cccc}
		\toprule Sr No. & Data Size & Avalanche Effect (New) & Avalanche Effect (Old) \\ \midrule
		1	& 49,152 & 50.04 \% & 34.43 \% \\ 
		2 & 709,200 & 50.05 \% & 31.11 \% \\ 
		3 & 889,200 & 50.37 \% & 33.81 \% \\ 
		4 & 950,878 & 50.11\% & 36.71  \%  \\ \bottomrule 
	\end{tabular}
\caption{Avalanche Effect of Improved Cipher}
\end{table}
We observe that PSNR and Entropy security parameters for improved cipher are almost identical to the old one and Avalanche effect is considerably increased to $50\%$, due to which it is relatively secure against related-key attack.
\section{Bio-SNOW: Newly Proposed DNA based Stream Cipher} \label{newcipher}

\subsection{Basic Design of Bio-SNOW}
There are two FSRs in the Stream Cipher denoted by $A$ an $B$ respectively. Both have size of 128 quads and labeled by $s$ and numbered from $0$ to $255$ i.e $s_0, s_1, s_2, \dots s_{255}$ where $s_0$ to $s_{127}$ are quads of LFSR-A and $s_{128}$ to $s_{255}$ are quads of LFSR-B. 

\par
Then, we have three FSMs(Finite State Machines) labelled as $R_1, R_2, $ and $R_3$. There are update mechanisms for these FSMs.
\begin{figure}[h]
	\begin{center}
		\includegraphics[scale=0.45]{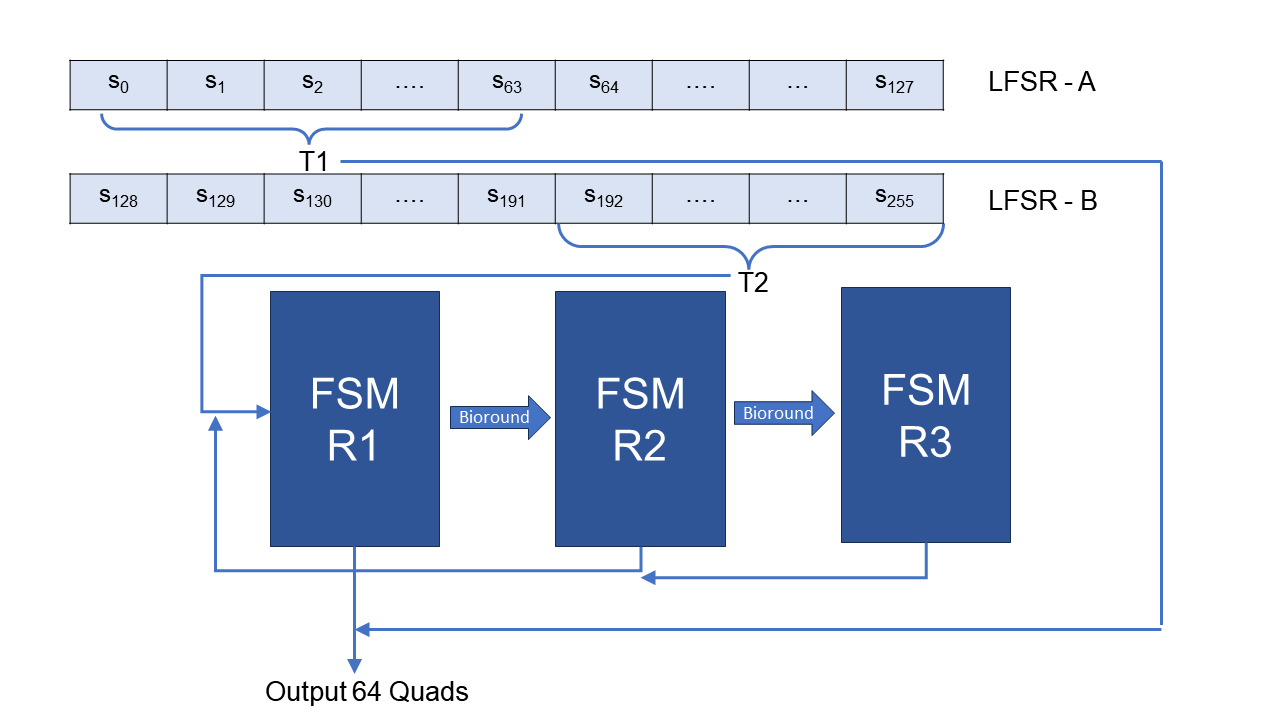}
		\caption{Structure of DNA based Stream Cipher}
	\end{center}
\end{figure}
\subsection{Cipher Algorithms:}
The operation of cipher can be given as:
\paragraph{Initialization Phase}
Let key $K=[K_1, K_2, \dots K_{127}] $ and Initialization vector be $IV= [IV_1, IV_2 \dots IV_{127}]$. \\
\begin{enumerate}
	\item \textbf{Loading of FSRs}We load the Key and IV to the FSR as: 
	\vspace{1cm}

	\begin{itemize}
		\item $[s_0, s_1, \dots s_{63}] = $ $ [K_{64}, K_{65}, \dots K_{127}]$.
		\item $[s_{64}, s_{65}, \dots s_{127}] =$  $ [IV_1, IV_2, \dots IV_{63}]$
		\item $[s_{128}, s_{129}, s_{130}, \dots s_{191}] =$  $ [IV_{64}, IV_{65}, \dots IV_{127}] $
		\item $[s_{192}, s_{193}, \dots s_{255}] = $ $ [K_{0}, K_1, \dots K_{63}]$
	\end{itemize}
	\item Update FSRs (4*256=1024) times. 
	\item Define Tap values as: 
	\[ T1= [s_{192}, s_{193}, \dots s_{255}]\]
	\[ T2 = [s_0, s_1, \dots s_{63}]\]
	\item for $i=1$ to $i=16$
	\begin{enumerate}
		\item Find $T_1$, $T_2$.
		\item Update FSRs $128$ times. 
		\item Update FSM one time. 
		\item If $t=15$, $R_1 = R_1 \oplus [K_0, K_1, \dots K_{63}]$
		\item If $t=16$, $R_1= R_1 \oplus [K_{64}, K_9, \dots K_{127}]$
	\end{enumerate}
\end{enumerate} 
The above steps complete the initialization phase, after which we can proceeed to Keystream generation phase. 
\paragraph{Keystream Generation}
To generate the Keystream, after initialization, we repeat the same steps and output the value \\
\begin{enumerate}
	\item Find $T_1$, $T_2$.
	\item Update FSRs $128$ times. 
	\item Update FSM one time. 
	\item  $Z= (R_1 \oplus T_1)$
\end{enumerate} \nopagebreak
This $Z$ is of size $64$ quads which is equivalent to $128$ bits and is used to BioXOR with the plaintext. \\
The encryption and decryption algorithm are same for symmetric ciphers.
\paragraph{Encryption and Decryption}
\begin{enumerate}
	\item Let $m$ be the plaintext as: 
	\[ m= m_0m_1m_2 \dots m_{2k}\]
	\item Encode $m$ into DNA string as : $00$ to A, $01$ to T, $10$ to C, and $11$ to G.
	\[ m= b_0b_1b_2 \dots b_k\] 
	\item Use BioXOR operation to encrypt the plain text into cipher text as: 
	\[ c_i=(b_i) \, \mbox{BioXOR} \, (z_i)\]   
\end{enumerate}

\subsection{Auxiliary Functions}
We have defined the Initialization and Keystream generation phase in the above subsection. Here we define the functions involved in the complete process. 

\begin{figure}[h]
	\begin{center}
		\label{fig:bioxormul}
		\includegraphics[scale=0.4]{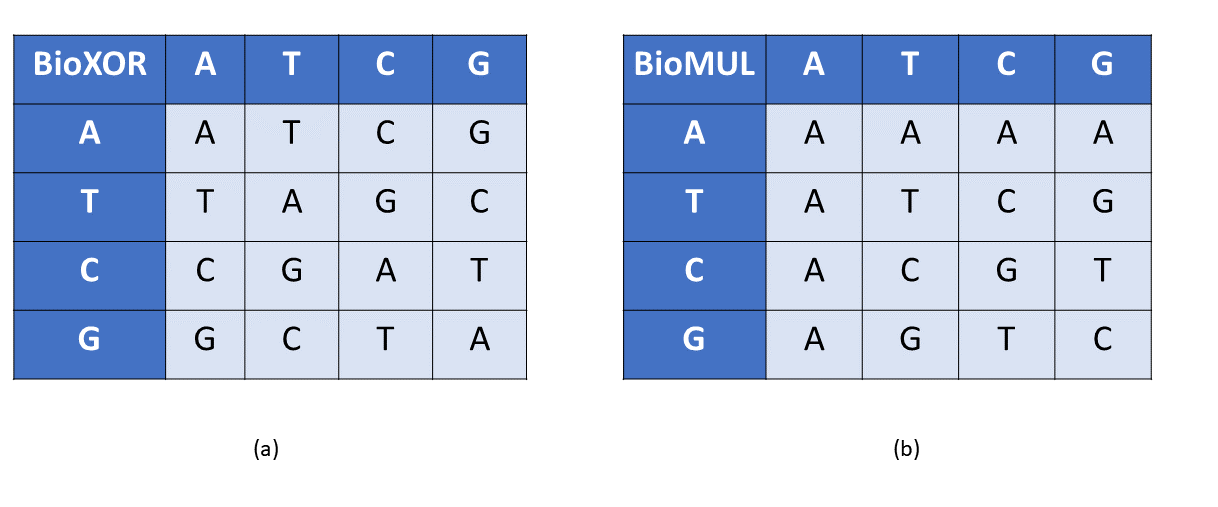}
		\caption{(a)BioXOR Table  (b) BioMUL Table}
	\end{center}
\end{figure}

\begin{enumerate}
	\item \textbf{BioXOR} We define the plus operation($\oplus$) in quad from Figure \ref{fig:bioxormul} (\cite{amin2006dna})
	
	\item \textbf{BioMul} The multiply operation($\otimes$) is defined from Figure \ref{fig:bioxormul}
	
	\item \textbf{FSR update}
	Given
	\[[s_0(k), s_1(k), s_2(k), \dots s_{255}(k)] \] as the state of FSRs at time $k$
	\begin{enumerate}
		\item $t_1= s_{100}(k) \oplus s_{127}(k) \oplus s_{126}(k)\otimes s_{125}(k) \oplus s_{249}(k)$
		\item $t_2= s_{240}(k) \oplus s_{255}(k) \oplus s_{253}(k) \otimes s_{254}(k) \oplus s_{114}(k)$
		
		\item $[s_0(k+1), s_1(k+1), \dots , s_{127}(k+1)] = [t_2, s_0(k), s_1(k), s_2(k), \dots, s_{126}(k)]$
		
		\item $[s_{128}(k+1), s_{129}(k+1), \dots , s_{256}(k+1)] = [t_1, s_{128}(k), s_{129}(k), \dots s_{255}(k)]$
	\end{enumerate}
	\item \textbf{ParallelAdd} Let $u$ and $v$ be two blocks of $64$ quads, then ParallelAdd($u, v$) is defined as first splitting $u$ and $v$ into blocks of $16$ quads and then adding them modulo $4$, taking carry to next element but not to the next $16-$ block. 
	\item \textbf{BioRound} For the BioRound function we have two sub-steps involved: 
	\begin{enumerate}
		\item \textbf{Transcription} In this process, substitution takes place which is inspired from the central dogma. We know that when DNA strand converts to mRNA, the changes take place as: 
		\[ [A, C, G, T] = [T, G, C, A]\]
		Actually $T$ is changed to $U$(Uracile), but we have defined as $T$ changing to $A$ for compatibility. \\
		So for an input $X$ of $64$ quads, each of $64$ quads is replaced according to above definition. 
		
		\item \textbf{Translation} This is also a simulation of process in central Dogma which we have referred by its own name Translation. For this we make collections of four consecutive quads in the input $X$. As input is size of $64$ quads, we get $16$ such sets. Then according to the Amino Acids Table given below, we substitute the set with the entry of the table in Figure \ref{fig:AminoAcid}
	\end{enumerate}
	
	This complete above process is defined as BioRound.
	\item \textbf{FSM update} 
	The updates of three FSM takes place as:

	\begin{itemize}
		\item $R1^{(t+1)}= $ ($R2^{(t)}$ \textbf{ParallelAdd} $R3^{(t)}$) $\oplus T2^{(t)}$
		\item $R2^{(t+1)}=$ \textbf{BioRound}($R1^{(t)}$)
		\item $R3^{(t+1)}=$ \textbf{BioRound}($R2^{(t)}$)
	\end{itemize}
\end{enumerate}

It generates keystream of $64$ quads which is equivalent to $128$ bits. 

\section{Speed and Security of Bio-SNOW}\label{securityandimage}
\subsection{Speed}
We compared the new proposed DNA Stream Cipher with the well known cipher used in mobile communication, the SNOW-3G. The source code of SNOW-3G is downloaded from  \href{https://github.com/mitshell/CryptoMobile.git}{github} which is a python wrapper over the C language. We developed code for Bio-SNOW in the python and there is a considerable difference in speed of C and python. To overcome this issue, we ran a same program in C and python and found that C language is around $32$ tims faster than Python.

\begin{table}[h]
	\centering
	\begin{tabular}{ccc}
		\toprule	Python & C &Ratio\\
		\midrule  40.568s &  1.226s & 33.08 \\ \bottomrule 
	\end{tabular}
\caption{Time taken by Python and C Language}

\end{table}

\vspace{0.2cm}

\textbf{Assumptions:} \\
\begin{itemize}
	\item Time taken is considered only for the Keystream generation phase with an average of $10$ observations. For comparison, the time taken by the python program of DNA stream cipher by 32 and then compare the results.
	\item The SNOW 3G source code takes argument $y$ as the number of $32-$ bit words whereas, our algorithm takes argument $x$ as the number of $64-$quad block or $128-$bit block. Therefore to maintain similarity in input we have to give input in such a way that: 
	\[ y=4x\]
	
\end{itemize}

The results for different pairs of given values of $x$ are: \\ 

\begin{table}[h]
	\centering
		\begin{tabular}[h]{lccp{4.5cm}}
			\toprule \textbf{Input size} & \textbf{Bio-SNOW} & \textbf{SNOW-3G Cipher} & \textbf{Times Bio-SNOW is faster than SNOW-3G.} \\ \midrule
			100 & 0.00058 & 0.0017 & 2.90 \\ 
			200 & 0.0012 & 0.00344 & 2.68 \\ 
			300 & 0.0018 & 0.0056 & 3.06 \\ 
			400 & 0.0024 & 0.0064 & 2.63 \\ 
			500 & 0.0029 & 0.0082 & 2.76 \\ 
			600 & 0.0035 & 0.0090 & 2.53 \\ 
			700 & 0.0041 & 0.011 & 2.66 \\ 
			800 & 0.0100 & 0.0252 & 2.51 \\ 
			900 & 0.0063 & 0.0164 & 2.48 \\ 
			1000 & 0.0065 & 0.0189 & 2.90 \\ 
			1100 & 0.0069 & 0.0175 & 2.53 \\
			1200 & 0.0078 & 0.0194 & 2.47 \\ \bottomrule 
		\end{tabular}
	\caption{Table of time taken by Bio-SNOW vs SNOW-3G}
\end{table}
From above we conclude that Bio-SNOW is around $2.5$ times faster than SNOW 3G Cipher. The graph of above data is given as: 
\begin{figure}[H]
	\begin{center}
		\includegraphics[scale=1.2]{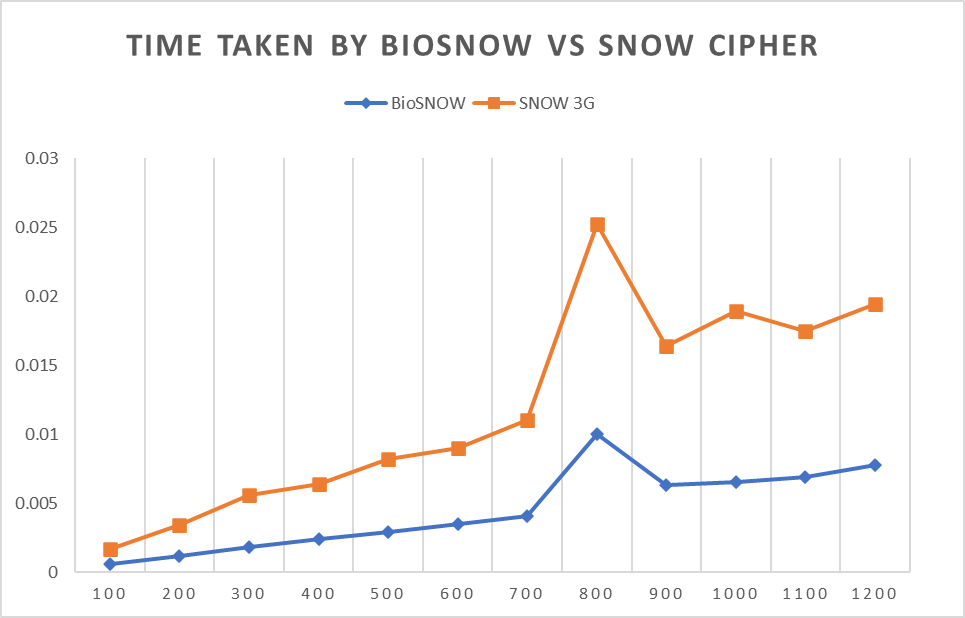}
	\end{center}
	\caption{Graph of comparison of time taken}
\end{figure}

\subsection{Security of BioSNOW}
\begin{enumerate}
	\item \textbf{Algebraic Attack:}
	We have used quads(A, C, G, T) over the bits. The attacker employing the attack has to solve equations in $F_4$ rather than $F_2$. Moreover, if he tries to build equations in $F_2$, it will be difficult as all operations are done under quads.
	The time complexity for equations over $F_2$ and $F_4$ is the same $O(n^3)$ but the operations will now be more computationally intensive. 
	\item \textbf{State-Recovery Attack:}
	To prevent a state recovery attack, we have used a well-known technique of key mixing by introducing key at step $15$ and $16$.
	\item \textbf{Correlation Attack:}
	Statistical attacks comprise the relation between the Feedback Shift Register and the output bitstream. To tackle such a situation, we use taps $T1$ and $T2$ from FSR to FSMs where the FSRs are updated $128$ times.
	\item \textbf{Non-Linearity:} We have used three FSMs for increasing the non-linearity in the bitstream.
	\item \textbf{Keyspace Analysis:} BioSNOW uses key size of $128$ quads and hence has keyspace of size $4^128= 2^256$. The large key space provides security against Brute Force attack. 
	\item \textbf{NIST Randomness Test:}
	The BioSNOW generates a stream of pseudorandom quads that are BioXORed with the plaintext. To be secure this stream must satisfy known randomness tests given by NIST(National Institute of Standards and Technology). We perform these tests on BioSNOW using a Python implementation of NIST's A Statistical Test Suite for Random and Pseudorandom Number Generators which can be found at \href{https://github.com/stevenang/randomness\_testsuite\#}{https://github.com/stevenang/randomness\_testsuite\#}. To strengthen our analysis, we tested five keystream samples generated from different keys, and all samples passed the tests successfully as shown in Fig \ref{NIST_test}
	
	\begin{figure}[h]
		\centering
		\includegraphics[scale=0.3]{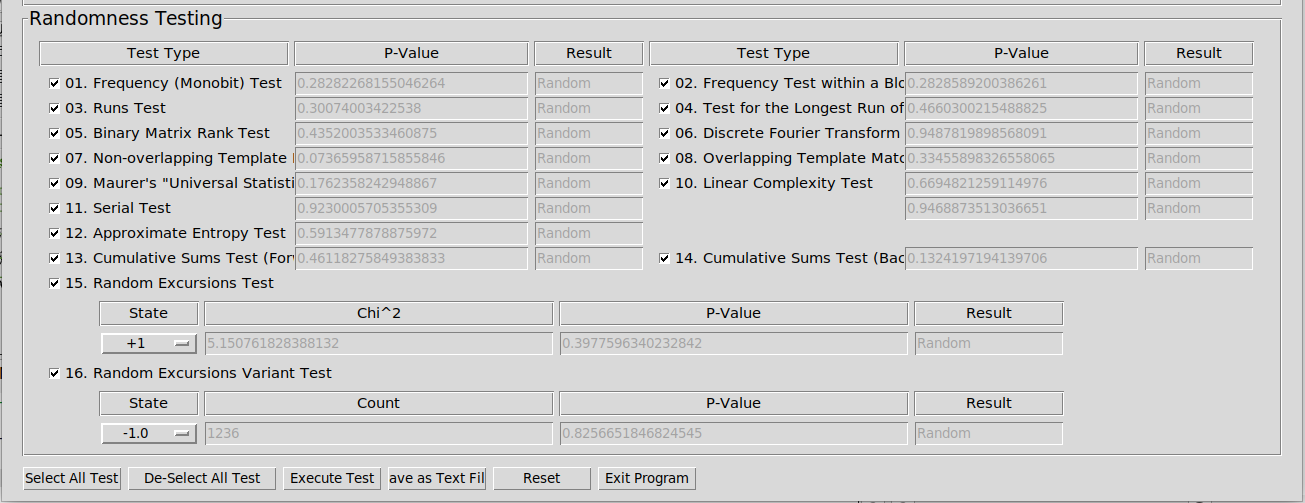}
		\caption{NIST Statistical Test}
		\label{NIST_test}
	\end{figure}

\end{enumerate}

\subsection{Application of Bio-SNOW in Image Cryptography}
The Bio-SNOW algorithm can also be used to encrypt images by XORing the pixel data of original image with the keystream. In case of raw data the encryption function of Bio-SNOW XORs the keystream with the plaintext data to generate the ciphertext. However, in this step if we encrypt an image, the ciphertext is not an image.

 We can extend the Bio-SNOW to act like image encryption algorithm. Let $p$ denote the original image. Split $p$  into $r,g,b$ parts named as $p_r, p_g, p_b$. let $(x,y)$ denote the pixel position, then $p_r(x,y)$ denote the intensity of red color pixel at position $(x,y)$. Similarly, we can define $p_g(x,y)$ and $p_b(x,y)$. Let $z$ denote the keystream generated by Bio-SNOW and $[z_i,\dots z_{i+3}]$ denote four consecutive quads. We get $c_r(x,y) = p_r(x,y) \oplus[z_i,\dots z_{i+3}]$. Taking consecutive quads to encrypt all color pixels so that no position of keystream is used twice in entrire process, we get three encrypted images $c_r, c_g,$ and $c_b$ that are combined to form the final encrypted image $c$. We can use this same algorithm for decryption of image given the same secret key $k$ for Bio-SNOW.

\begin{figure}[H]
	\centering
	\begin{minipage}{0.3\textwidth}
		\centering
		\includegraphics[width=\linewidth]{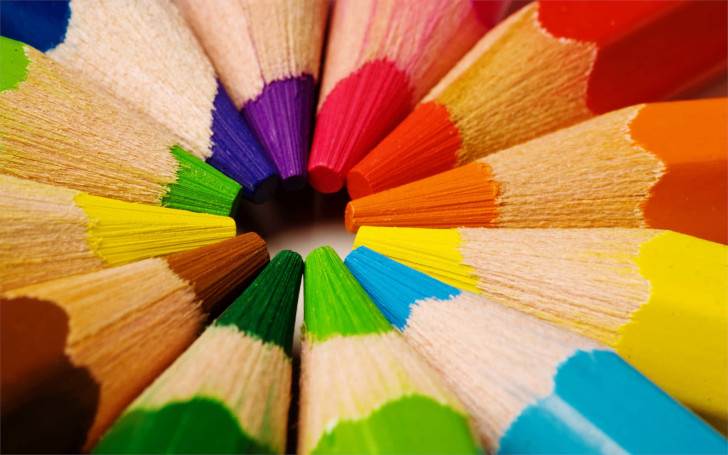}
		\caption{Original Image}
		\label{fig:image1}
		\includegraphics[width=\linewidth]{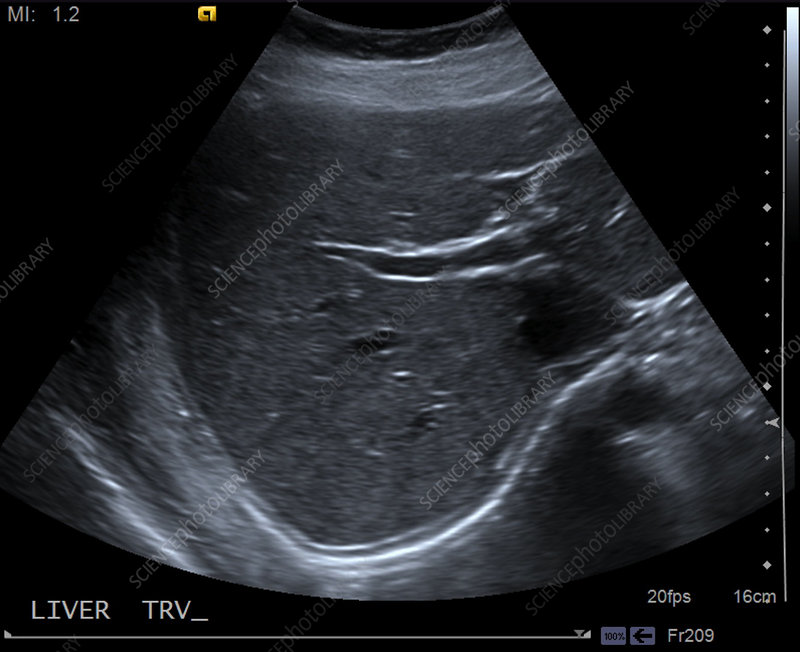}
		\caption{Original Medical Image}
		\includegraphics[width=\linewidth]{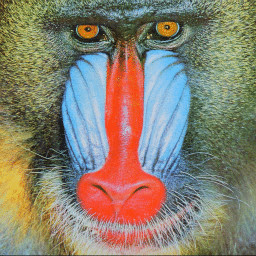}
		\caption{Original Baboon Image.}

	\end{minipage}
	\hspace{0.03\textwidth} 
	\begin{minipage}{0.3\textwidth}
		\centering
		\includegraphics[width=\linewidth]{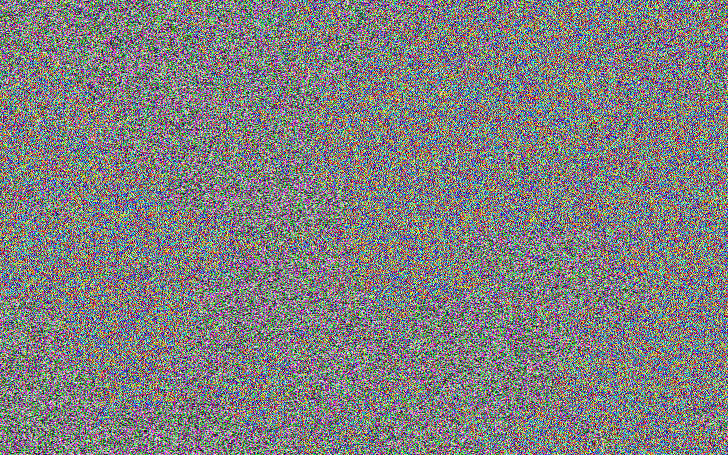}
		\caption{Encrypted Image}
		
	\includegraphics[width=\linewidth]{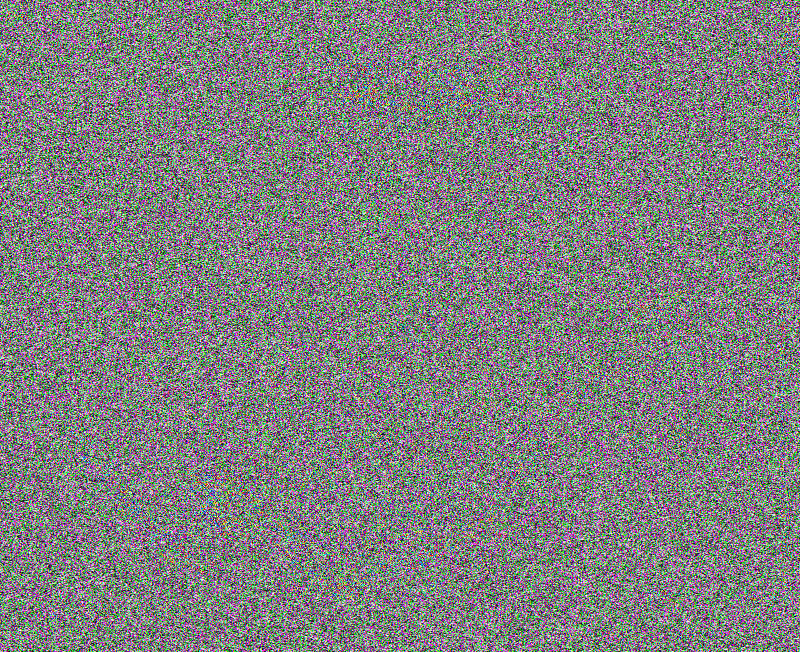}
	\caption{Encrypted Medical Image}
		\label{fig:image2}
	\includegraphics[width=\linewidth]{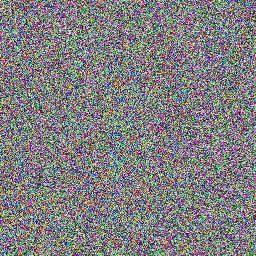}
	\caption{Encrypted Baboon  Image}
	\end{minipage}
	\hspace{0.03\textwidth} 
	\begin{minipage}{0.3\textwidth}
		\centering
		\includegraphics[width=\linewidth]{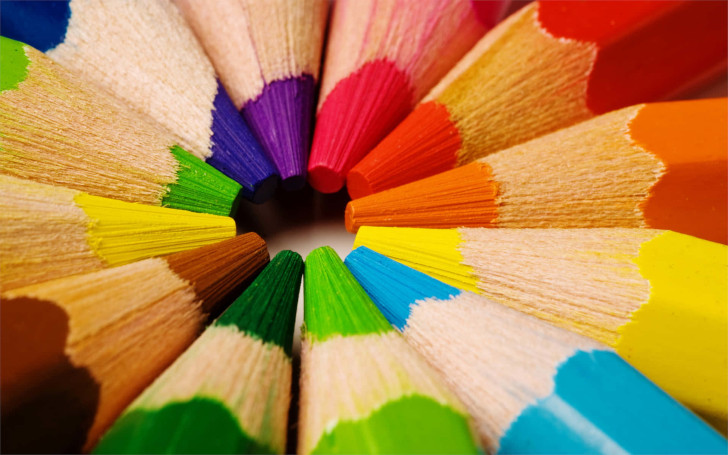}
		\caption{Decrypted Image}
			\includegraphics[width=\linewidth]{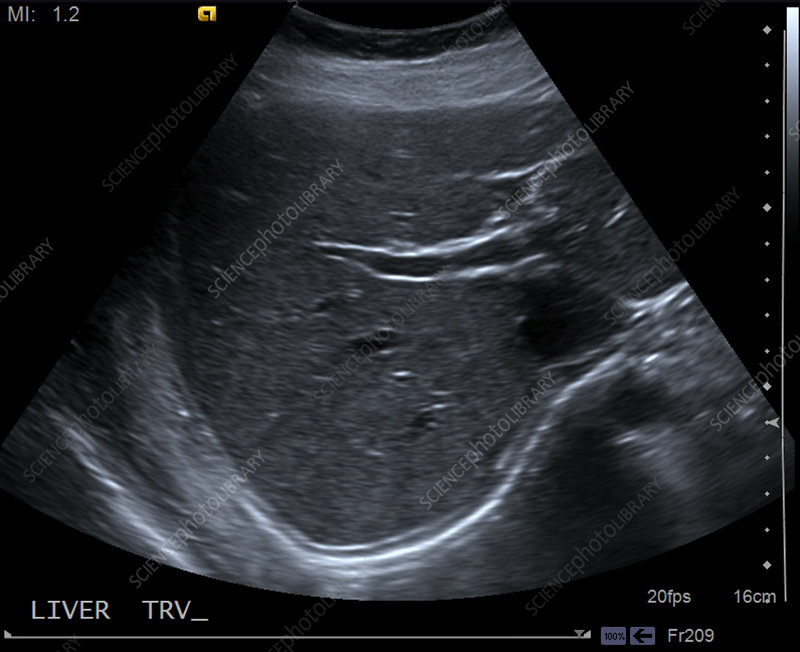}
			\caption{Decrypted Medical Image}
		\includegraphics[width=\linewidth]{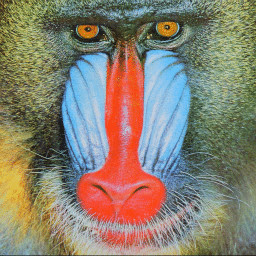}
		\caption{Decrypted Baboon Image}
		\label{fig:image3}
	\end{minipage}
	\label{fig:combined}
\end{figure}

\subsection{Security Analysis of Image Cryptography}
\subsubsection{Histogram Analysis}

Histogram analysis shows the spread of pixel values over the image. We first split image into RGB formats and then count the number of pixels in image having a fixed intensity value. For eg. Considering red part of the image, we count the number of pixels for $i=0$ to $i=255$ and plot them on a graph. The existence of peaks in histogram shows non-randomness and uniformity depicts the randomness in image. 
\begin{figure}[H]
	\centering
	\includegraphics[scale=0.35]{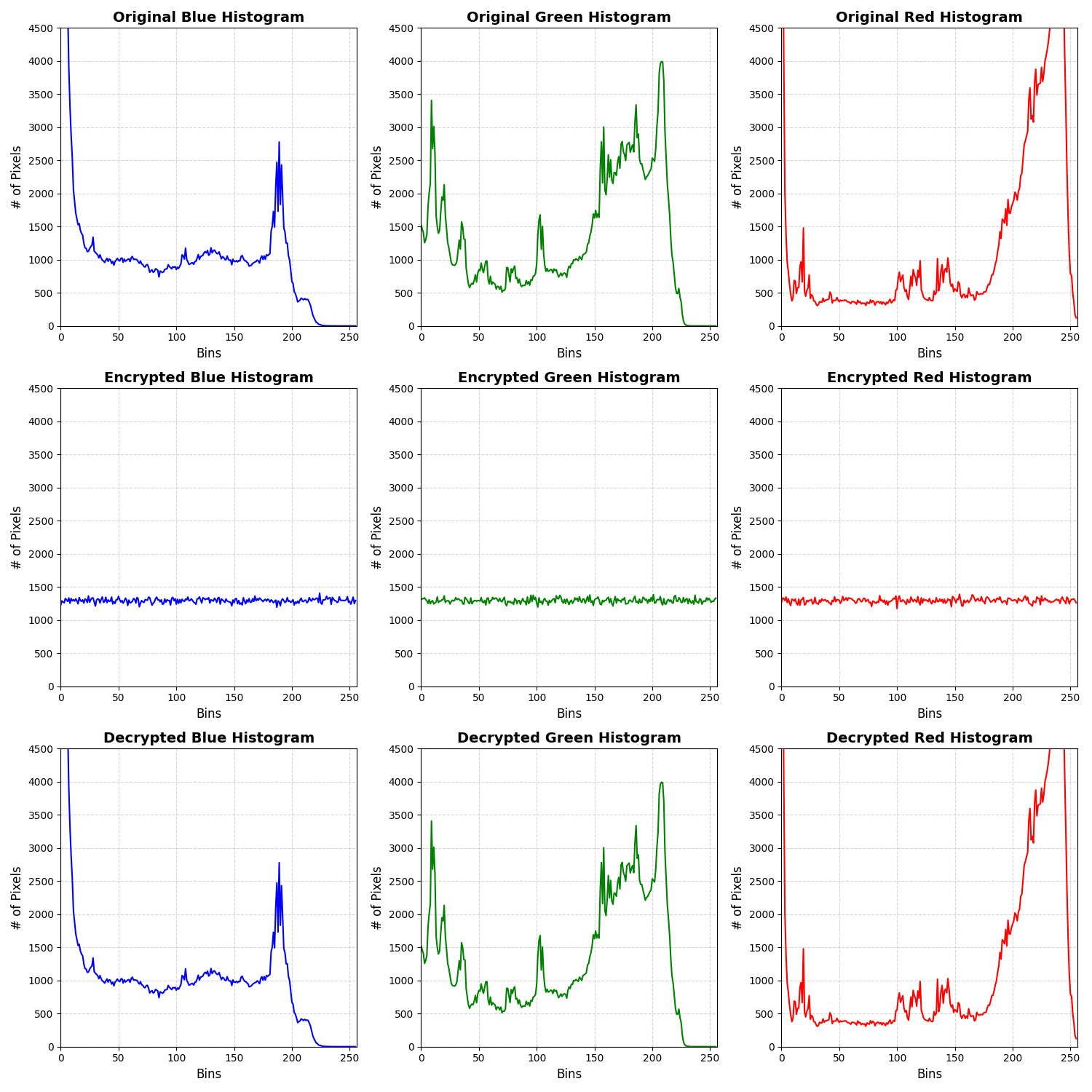}
	\caption{Histogram comparison of original, encrypted and decrypted image}
\end{figure}
\vspace{1cm}

\subsubsection{Correlation Analysis}
Correlation analysis of image encryption is a crucial method used to evaluate the effectiveness of encryption algorithms. This analysis examines the statistical relationship between adjacent pixels in an image. 
In the context of image encryption, correlation coefficients are calculated for pixel pairs in horizontal, vertical, and diagonal directions for each color channel (red, green, and blue). These coefficients are then compared for the original, encrypted, and decrypted images in Table \ref{cortable}. For the original image, adjacent pixels tend to have high correlation coefficients, indicating that the pixel values are closely related and visually similar. Effective encryption should significantly reduce this correlation, making the pixel values appear random and independent. 

\begin{table}[h] 
\begin{tabular}{lllll}
	\toprule
	&   & Original & Encrypted  & Decrypted  \\
	\midrule 
	
	Blue & Horizontal &0.993637 &0.001466141 &0.9936370 \\
	& Vertical &0.987257 &-0.0018053783 & 0.987257\\
	& Diagonal & 0.983986&-0.002179703 &0.983986 \\
	Green & Horizontal &0.995206&0.000641153&0.995206 \\
	& Vertical &0.991687 & -0.001249218& 0.991687\\
	& Diagonal &0.987632 & 0.0004761345&0.987632 \\
	Red & Horizontal &0.995408&0.000695197&0.995408 \\
	& Vertical &0.993864 &0.001263818 & 0.993864\\
	& Diagonal &0.990576 & 0.000878260&0.990576 \\	
	\bottomrule
\end{tabular}
\caption{Pearson's Correlation Coefficient of Original, Encrypted and Decrypted Images}
\label{cortable}
\end{table}
To illustrate this, scatter plots are often used to depict the relationship between pixel values of adjacent pairs in the three directions for each color channel. These plots visually represent how the encryption process disrupts the predictable relationships between adjacent pixels.

By comparing these correlation coefficients and scatter plots for the original, encrypted, and decrypted images, researchers can quantitatively assess the robustness of the encryption algorithm. This comprehensive correlation analysis provides a detailed mathematical insight into the quality and security of image encryption methods.

 We found the following scattering plot regarding correlation for green part of encrypted image. The blue and red also give similar results and are omitted. It suggests that our encryption function doesn't have any statistical relationships and hence can be considered a secure cipher.  \\

\begin{figure}[H]
	\centering
	\includegraphics[scale=0.28]{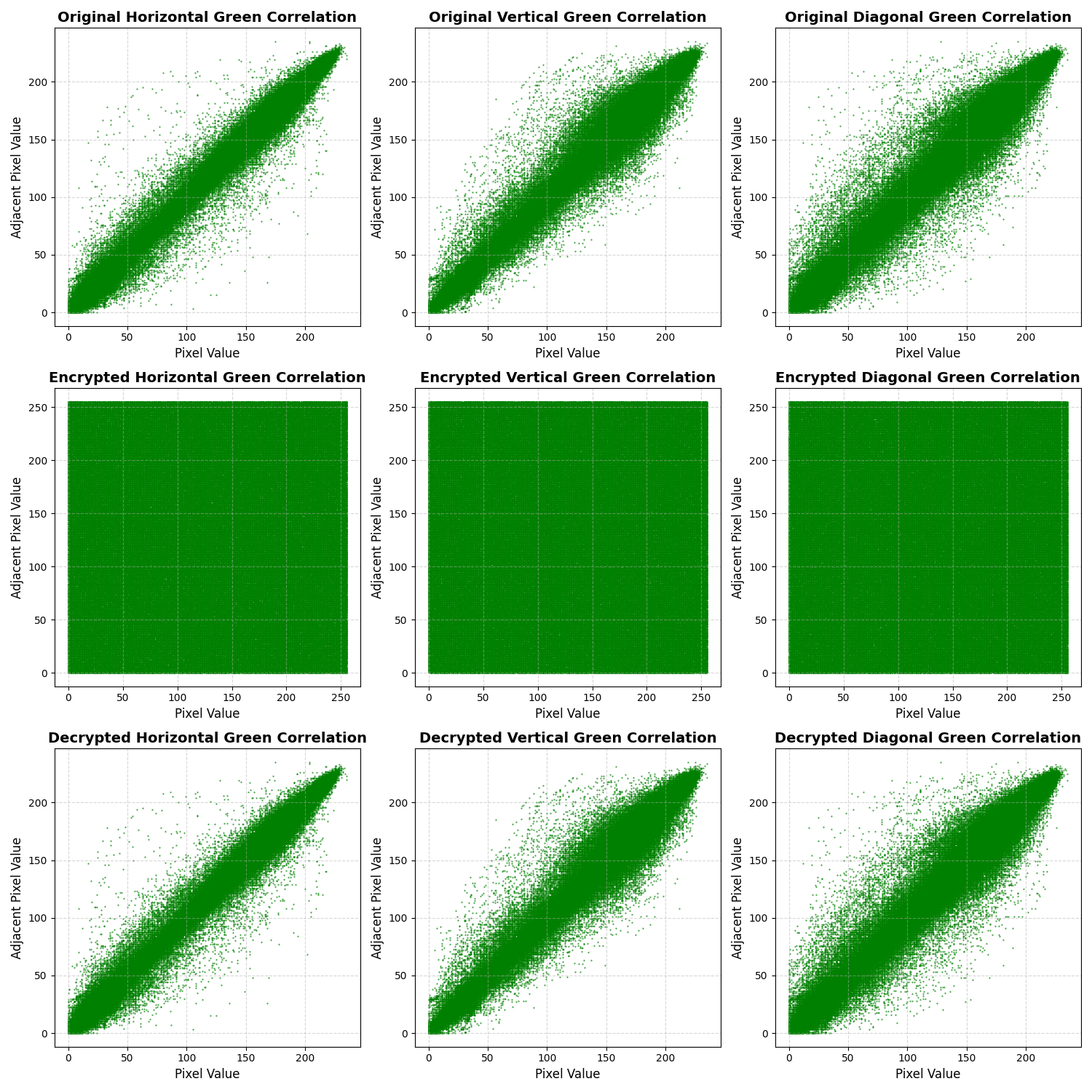}
\end{figure}

In a secure encryption scheme, \( r_{\text{horizontal}} \), \( r_{\text{vertical}} \), and \( r_{\text{diagonal}} \) should approach zero, indicating effective decorrelation of adjacent pixels, thereby enhancing resistance to statistical attacks. \\ In our analysis, we observed that the correlation coefficients for the cipher image of the Baboon have been reduced to nearly zero. This significant reduction in correlation indicates effective pixel decorrelation achieved by the encryption algorithm. To evaluate the performance of our method, we compared these coefficients with those reported in recent literature in Table \ref{corliterature}

\begin{table}[h]  
\begin{tabular}{lccc}
\toprule
Algorithm & Horizontal &Vertical & Diagonal \\
 \midrule Proposed &  0.01431 & -0.00094& -0.00132 \\
\cite{GUO202449} & 0.0042 & 0.0049 & 0.0050 \\
\cite{LIANG2023109033}&  0.0120 & 0.0091 & 0.0056 \\ \cite{YE2022117709}&0.0161 & 0.0063 &  0.0170 \\ \bottomrule 
\end{tabular}
\caption{Comparison of correlation coefficient in grayscale with recent literature}
\label{corliterature}
\end{table}

\section{Conclusion and Future Direction}\label{sec:conclusion}
In conclusion, our examination of a DNA Cipher for lightweight cryptography highlighted multiple insecurities. Specifically, our analysis identified vulnerabilities in the form of related key, Distinguisher, key reduction, and complete break attacks. The system's reliance on a linear equation system with just $2n$ variables allows for swift resolution within fractions of a second, exposing a fundamental weakness. Moreover, a weaker Avalanche Effect paved a way for Distinguishing and Related Key Attacks. We proposed a straightforward enhancement to mitigate this particular vulnerability using AminoAcid S-Box and mutator vector in the encryption process. Nonetheless, it is crucial to note that additional vulnerabilities might exist, potentially susceptible to different types of attacks. Investigating these potential vulnerabilities represents a promising avenue for future research in this multidisciplinary field linking Biology and Cryptography.

Subsequently, our study presents a novel DNA-based stream cipher---Bio-SNOW--- that demonstrates a notable performance advantage, being approximately $2.5$ times faster than the SNOW 3G cipher. This comparison accounts for various external factors, including programming language and output data size. Moreover, in assessing security aspects, our findings indicate enhanced results achieved by augmenting the size of Finite State Machines (FSMs) and leveraging inherent natural randomness from DNA, among other contributing factors. This cipher may be used in wearable and implantable devices such as fitness trackers, glucose monitors, Pulse Oximeters and  cardiac pacemakers, Cohecular implants respectively.
\par 
The experimental results, including histogram and correlation analyses, validate the efficacy of Bio-SNOW in image cryptography. The histogram analysis demonstrates that the encrypted images exhibit a uniform distribution of pixel values, while the correlation analysis confirms the successful disruption of pixel dependencies. These results affirm that cipher Bio-SNOW provides a robust and secure method for image encryption, maintaining the cryptographic strength essential for secure image transmission and storage.
\par 

In the near future, researchers have a promising opportunity to delve into uncovering weaknesses in both the Enhanced DNA-based Lightweight Cipher and the Bio-SNOW. Furthermore, they can focus on optimizing these ciphers, akin to the advancements seen with the SNOW cipher in communication technology.
\section{Funding}
The First author was supported by the University Grants Commission Junior Research Fellowship (UGC-JRF) with Reference No. 211610136084.
\section{Data Availability Statement}
The data that support the findings of this study, as well as the source code of the programs used in the analysis, are available from the corresponding author upon reasonable request.
\section{Declaration of Interests}

The authors declare that they have no known competing financial interests or personal relationships that could have appeared to influence the work reported in this paper.

\bibliography{biblio}

\end{document}